\documentclass{acm_proc_article-sp}

\makeatletter
\newif\if@restonecol
\makeatother

\usepackage[boxruled,vlined]{algorithm2e}
\usepackage{amssymb,amsmath}
\usepackage[utf8]{inputenc}
\usepackage{graphicx}
\usepackage{color}
\usepackage{times}
\usepackage{xspace}
\usepackage[textsize=scriptsize,disable]{todonotes}
\usepackage{appendix}
\usepackage{url}
\usepackage{flushend}
\usepackage{subfigure}
\usepackage{url}
\usepackage{tabularx}
\newcolumntype{Y}{>{\raggedright\arraybackslash}X}
\usepackage{booktabs}

\newcommand{\plotwidth}{0.45}

\begin{document}

\title{The Impact of the Filter Bubble---A Simulation Based Framework for Measuring Personalisation Macro Effects in Online Communities}
\numberofauthors{2}
\author{
\alignauthor
Thomas Gottron\\
       \affaddr{Institute for Web Science and Technologies,\\University of Koblenz-Landau, Germany}\\
       \email{gottron@uni-koblenz.de}
\alignauthor
Felix Schwagereit\\
       \affaddr{Institute for Web Science and Technologies,\\University of Koblenz-Landau, Germany}\\
       \email{schwagereit@uni-koblenz.de}
}
\maketitle

\begin{abstract}
The term \emph{filter bubble} has been coined to describe the situation of online users which---due to filtering algorithms---live in a personalised information universe biased towards their own interests.
In this paper we use an agent-based simulation framework to measure the actual risk and impact of filter bubble effects occurring in online communities due to content or author based personalisation algorithms.
Observing the strength of filter bubble effects allows for opposing the bene\-fits to the risks of personalisation.
In our simulation we observed, that filter bubble effects occur as soon as users indicate preferences towards certain topics.
We also saw, that well connected users are affected much stronger than average or poorly connected users.
Finally, our experimental setting indicated that the employed personalisation algorithm based on content features seems to bear a lower risk of filter bubble effects than one  performing personalisation based on authors.
\end{abstract}

%%%%%%%%%%%%%%%%%%%%%%%%%%%%%%%%%%%%%%%%%%%%%%%%%%%%%%%%%%%%%%%%%%%%%%%%%%%%%%%%
\section{Introduction}
\label{sec:introduction}
%%%%%%%%%%%%%%%%%%%%%%%%%%%%%%%%%%%%%%%%%%%%%%%%%%%%%%%%%%%%%%%%%%%%%%%%%%%%%%%%

Eli Pariser coined the term filter bubble in~\cite{Book:2011:Pariser}.
The filter bubble represents the personal, unique universe of online information a user lives in.
The boundaries of this universe, what gets in and what remains outside the reach of the user, says Pariser, depends on the personalisation algorithms in social networks and search engines.
The personalisation algorithms consider a user's interaction with information to predict which information she most likely considers relevant.
As a consequence the algorithms will show such information at higher ranks in a result list more likely or, in the most extreme case, even exclusively, i.e. blocking other information from the sight of the user.
Thus, the personalisation algorithms effectively operate as filters on the information the users perceive.

The genuine aim of the personalisation algorithms is to get an idea of the topics the users are actually interested in and to improve their user experience by preferably showing them contents that match their interests.
On the downside, this leads to a situation, where the user might be confronted with information that is so highly personalised, that she will not see any content that does not match her interests or represents opinions and facts apart from her own point of view.
For an individual user this might lead to a skewed and biased perception of the world.

The risks and problems of the filter bubble effect have been discussed widely and in different contexts. 
Several approaches aim at ensuring diversified results in search scenarios or for particular information needs~\cite{drosou2010search}. %\todo{other diversification papers here} 
The actual risk of observing filter bubble effects as well as their strength, instead, has hardly been analysed.

In this paper we use an agent-based simulation to measure the impact of personalisation algorithms on the perception of users in social networks.
Here, we are not interested in the effects on the level of individual users, but rather on the macro level of an entire community.
To this end, we combine an established generative network model~\cite{J:Science:1999:BarabasiA} and a state of the art topic modelling approach~\cite{J:MLR:2003:BleiNJ} to simulate a social network structure and the messages the users create and consume in this network. 
The resulting model consists of a set of agents, each of which with her individual range of topics she is interested in.
The agents have their individual network context of friends.
They send messages to and receive messages from these friends.
By involving algorithms for personalisation the simulation allows to observe the development of appropriate macro-level metrics which measure the benefit and risk of personalised filters.
The metrics to measure the benefits include classical information retrieval metrics, such as precision and recall.
The negative impact is measured by analysing the reduction of the active social context to only a few friends or a reduced used of vocabulary.
Finally, the number of perceived messages that fall into the core field of interest of the users can be read in both ways: positive as the users are not bothered with off-topic messages and negative because their perception of other topics of peripheral interest is reduced or even filtered out entirely.

We considered two possible algorithms for personalisation: one based on the content of messages and one based on the author of messages.
In our simulation we observed, that both personalisation approaches lead to filter bubble effects w.r.t. the focus of the perceived messages being shifted towards the core interests of the users.
However, we also observed that the personalisation algorithm based on content lead to the same benefits, but with a lower impact on the downsides, e.g. the active social network does not become as thin, neither the used vocabulary as sparse.

The contributions we make in this paper are the following:

\begin{itemize}

\item We describe a model for simulating the generation, dissemination and rating of messages based on established generative models for social network construction and topic models.

\item We investigate the strength of filter bubble effects on a macro level when introducing personalisation filters in the content dissemination of a social network.

\item We analyse which user groups of an online community are affected most by filter bubble effects.

\item We compare two prototypical filtering methods under the aspect of risks and benefits of filter bubble effects.

\end{itemize}

We proceed as follows.
In Section~\ref{sec:relWork} we will review related work in the field of community simulation as well as the analysis of effects of personalisation and diversification in information provision.
Afterwards we present the general framework of our simulation.
Its implementation and the choice of model parameters is presented in~\ref{sec:implementation}.
In Section~\ref{sec:personalisation} we describe the two personalisation algorithms we have analysed and in Section~~\ref{sec:metrics} the metrics we use to measure filter bubble effects.
The actual experiments in the form of simulation runs are discussed in Section~\ref{sec:experiments}.
We present our results in~\ref{sec:results} and discuss them in detail in~\ref{sec:discussion}.
We end in Section~\ref{sec:summary} with a summary of our findings and a look at future work.

%%%%%%%%%%%%%%%%%%%%%%%%%%%%%%%%%%%%%%%%%%%%%%%%%%%%%%%%%%%%%%%%%%%%%%%%%%%%%%%%
\section{Related Work}
\label{sec:relWork}
%%%%%%%%%%%%%%%%%%%%%%%%%%%%%%%%%%%%%%%%%%%%%%%%%%%%%%%%%%%%%%%%%%%%%%%%%%%%%%%%

The importance and the effects of information propagation and perception in social networks and information portals has been analysed in various contexts.
Serendipity discovery of non-relevant items in digital libraries~\cite{P:ECDL:2009:TomsM} shows a general benefit of being inspiring for seeking information and thinking out of the box.
The impact on political processes and democracy was discussed in~\cite{P:PLEAD:2012:Barocas}.
A psychological study addressed the effects of information presentation and forgetting processes in social networks~\cite{J:JEP:2012:ComanH}.

Simulation and probabilistic models as tools for the analysis of online communities has been established in recent years.
Schwagereit et al. use simulations to support policy modelling and strategic decisions~\cite{P:WebSci:2010:SchwagereitSS}.
The dynamics of popularity among news stories was investigated and described by Hogg and Lerman~\cite{P:ICWSM:2010:HoggL,J:TIST:2012:LermanH}.
Self-enforcing content generation mechanisms and their representation as stochastic processes were investigated in~\cite{P:EC:2008:Wilkinson}.

In our simulation, we use the preferential attachment model of Barabasi and Alberts~\cite{J:Science:1999:BarabasiA} to construct a social network.
Preferential attachment is a probabilistic model to generate graph structures which exhibit typical features of social networks.
We use topic models to represent the interests of agents.
Latent-Dirichlet-Allocation (LDA)~\cite{J:MLR:2003:BleiNJ} can be considered a state of the art topic modelling approach.
LDA is a generative, probabilistic model to describe the topic composition of text documents.
To this end each document is represented as distribution over $k$ global topics.
This distribution is independent from the distribution of other documents but follows the same parameters.
The topics themselves are modelled as distributions over terms.
Various extensions have shown, that LDA is also suitable to model sentiments and opinions as well as authors.

To the best of our knowledge there is no systematic large scale analysis of filter bubble effects, neither based on real world data nor based on simulations.

%%%%%%%%%%%%%%%%%%%%%%%%%%%%%%%%%%%%%%%%%%%%%%%%%%%%%%%%%%%%%%%%%%%%%%%%%%%%%%%%
\section{The Simulation Model}
\label{sec:simulation}
%%%%%%%%%%%%%%%%%%%%%%%%%%%%%%%%%%%%%%%%%%%%%%%%%%%%%%%%%%%%%%%%%%%%%%%%%%%%%%%%

For our analysis we make use of an agent-based simulation.
The initial step prior to starting the actual simulation consists of the setup of the network with a fixed number of agents, each of which has a profile of her individual interests.
The network structure reflects the friendship relations between agents and will provide the basis for how  messages are disseminated among the agents.
We do not consider changes in the network structure during the simulation of the communities interaction.
An agent's profile of interest will serve to describe the messages she will write as well as for the decision whether or not a message lies within her core field of interests.

\begin{figure}
\centering
    \includegraphics[width=0.4\textwidth]{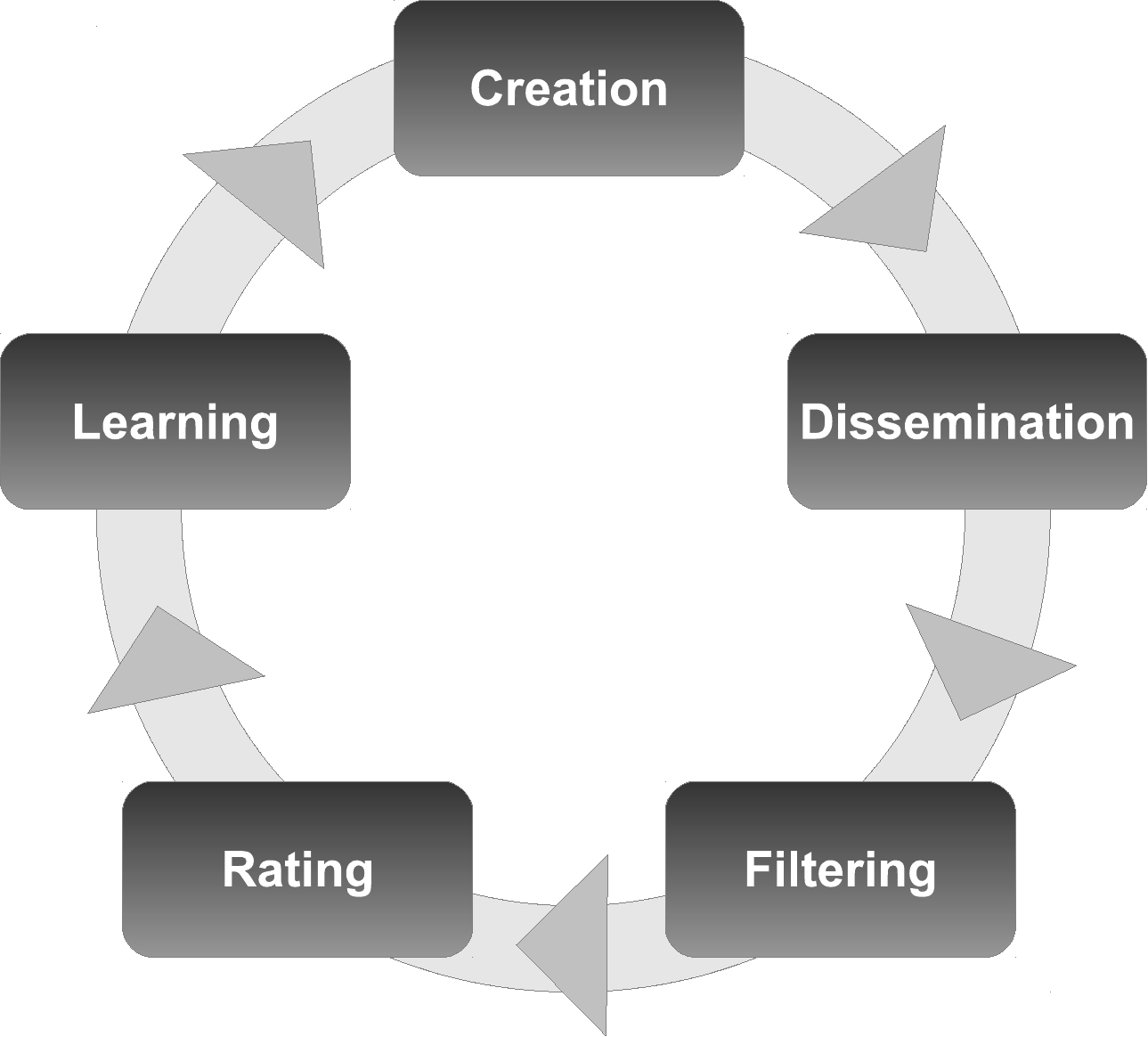} 
\caption{High level iterative process of the agent-based simulation.}
\label{fig:process}
\end{figure}

The idea of the simulation is focussing on an iterative creation, dissemination, filtering and rating of messages published in a social network plus a step of learning for updating the filters based on the user feedback.
The overall process is depicted in Figure~\ref{fig:process} and we illustrate the high level steps before going into the technical details of the implementation in the next section.

\begin{description}
\item[Creation] Each agent creates messages. 
The number of messages an agent creates depends on how verbose she is.
The content of the messages themselves depends on an agent's interests.

\item[Dissemination] The messages are disseminated along the edges of the social network. 
This means that for each agent the set of incoming messages is determined by the messages created by her friends.

\item[Filtering] The personalisation algorithms operate on this set of incoming messages.
The messages are ranked according to the predicted relevance and the result list is cut off at a given threshold.
Each agent has an individual instance of the personalisation algorithm which is specifically trained on the basis of her behaviour.

\item[Rating] The agents are presented with the messages the personalisation algorithms deemed interesting for them.
At this point the agents interact with the messages of other agents, i.e. they rate them as relevant or as irrelevant.
This decision also depends on whether or not the message is in the core field of interest of a agent.

\item[Learning] The personalisation algorithms receive the information which of the messages were rated as relevant.
This feedback of the agents is used by the personalisation algorithms to update their individual relevance model for the agents.

\end{description}

Iterating this cycle many times allows the personalisation algorithms to learn from the agent's feedback.
In this way they can adapt to the agent's interests.
At some point in time we would expect to see filter bubble effects, such as a high ratio of messages in the core field of interest, the active social network of the agents being reduced to a few contacts or a reduction of the perceived vocabulary.

%*******************************************************************************
\section{Implementation}
\label{sec:implementation}
%*******************************************************************************

The general process of the simulation is given as pseudo code in Algorithm~\ref{algo:simulation}.
The phase of the initial setup and the iteration phase with the steps of content generation, dissemination, filtering, rating and the learning phase are outlined.
The next sections illustrate the relevant parts of this process, i.e. the concrete implementation of the methods in this algorithms.

\begin{algorithm}
    \caption{Simulation}
    \label{algo:simulation}

    \KwIn{Number of agents $N$, personalisation algorithm $F$, Number $I$ of iterations}
    \tcp{Agent Setup}
    \For{$n = 1 \ldots N$}{
       \tcp{Create an agent and her instance of the personalisation algorithm}
       $a \leftarrow \textsc{initialiseAgent}()$\;
       $F_{a} \leftarrow F.\textsc{instance}()$\;
       $A \leftarrow A \cup \{a\}$;
    }
    \tcp{Network Setup}
    $\textit{Network} \leftarrow \textsc{constructNetwork}()$\;
    \For{$i = 1 \ldots I$}{
       \tcp{Content creation}
       \ForEach{$a \in A$}{
          $\textit{out}_{a} \leftarrow a.\textsc{generateMessages}()$;
       }
       \tcp{Dissemination}
       \ForEach{$a \in A$}{
          \ForEach{$b \in \textit{Network}.\textsc{friendsOf}(a)$}{
             $\textit{in}_{a} \leftarrow \textit{in}_{a} \cup \textit{out}_{b}$;
          }
       }
       \tcp{Filtering}
       \ForEach{$a \in A$}{
          $\textit{filtered}_{a} \leftarrow F_{a}.\textsc{filter}(\textit{in}_{a})$;
       }
       \tcp{Rating}
       \ForEach{$a \in A$}{
          \ForEach{$c \in \textit{filtered}_{a}$}{
             \If{ $a.\textsc{isRelevant}(c)$} {
                $\textit{relevant}_{a} \leftarrow \textit{filtered}_{a}	\cup \{c\}$;
             } 
          }
       }
       \tcp{Learning}
       \ForEach{$a \in A$}{
          $F_{a}.\textsc{learnInterests}(\textit{filtered}_{a}, \textit{relevant}_{a})$;
       }
    }
\end{algorithm}

\subsection{Initialising an Agent and her Interests}

When initialising an agent with the method \textsc{initialiseAgent}, the interests of the agents are modelled in a probabilistic manner by randomly assigning a distribution over topics.
The more probable topics in this distribution represent the core field of interest of the agent.
This distribution will be used in both the generative process when an agent writes messages as well as in the perception phase when an agent decides whether or not a message lies within her core field of interests.

To implement the topic models of the agents we employ Latent Dirichlet Allocation (LDA)~\cite{J:MLR:2003:BleiNJ}.
To this end, each agent is assigned a random distribution over $k$ global topics which is drawn from a Dirichlet distribution with prior $\alpha$.
The entries in the prior vector $\alpha$ influence how the distribution over the topics looks like. 
High entries in $\alpha$ generate distributions in which more topics are covered equally likely.
Low values in $\alpha$ generate topic distributions in which only few topics have a high probability of occurring.
The global topics themselves are randomly initialised as well.
They also follow a Dirichlet distribution.
Here the prior $\beta$ influences how the distributions of terms within each topic look like. 
Thus, the topic distribution of an agent is independent from the distribution of the other agents but follows the same parameters.
The same is true for the topic distributions.
As we are not interested in the actual contents, we use an artificial vocabulary and artificial topics, which have the same probabilistic characteristics as real world topics.
We chose the values of $\alpha$ and $\beta$ from literature to model focussed topics and a not too wide coverage of too many topics by an individual agent.
The concrete settings for the Dirichlet priors, as well as the number of agents and vocabulary terms we used in our simulation runs are summarised in Table~\ref{tab:parameters}.

Additionally we model the verbosity of each agent individually.
To this end, we provide each agent $a$ with the number $\lambda_a$ of messages we expect her to produce on average.
The value of $\lambda_a$ is chosen randomly when setting up the agent and follows a $\chi$ distribution with mean $e$.
To have realistic values for how many messages we can expect the user to write 
we estimated the value of $e$ from statistics on how many tweets people write per day\footnote{\url{http://edudemic.com/2012/12/14-twitter-statistics-you-may-not-know/}}.

\subsection{Social Network Setup}

We construct the social network of the agents randomly in \textsc{constructNetwork} using the preferential attachment model~\cite{J:Science:1999:BarabasiA}. 
In this generative model a social network graph is constructed starting from a nucleus of very few completely connected nodes.
Iteratively the graph is extended with new nodes, each of which establishing the same amount of edges to existing nodes.
The probability of connecting to an existing edge is proportional to the number of edges it already has.
In our settings, we started with a nucleus of ten nodes and had each newly created node connect to five other nodes.

\subsection{Content Creation}

The first step in the method \textsc{generateMessages} is that the agent randomly decides how many messages she is going to write in this iteration.
The number of generated messages follows a Laplace distribution with parameter $\lambda_a$, which has been assigned to each agent $a$ during initialization.

Once it is decided how many messages will be generated, the agents write each individual message $c$ based on their profile of interests.
For the sake of simplicity, in our simulation each message has an equal length of 10 words.
To determine the actual words we used the generative process behind the LDA model.
This means that for each word of the messages, the model randomly chooses first a topic (according to the agent's topic distribution) and in a second step randomly selects a word from this topic (according to the term distribution of the corresponding topic).

\subsection{Rating}

The agents need to be able to interact with the messages in the social network in order to give feedback on what is relevant to them.
To this end, in the method \textsc{isRelevant} each message the agent perceives is first classified into either falling into her field of core interests or into the category of peripheral interest.
The core interest of an agent is defined by the set of LDA topics which covers a high fraction of her individual topic distribution.
The other topics describe her peripheral interests.
The decision if a message is in the field of core interests is based on the odds of the agent to have constructed such a message solely from her core interests topics versus solely from her peripheral interests.
If these odds are higher than a given threshold $\theta$, a message is considered to be of core interest to an agent, otherwise it belongs to her peripheral interests.
For our simulation we set the value \emph{cov} of core topic coverage to 80\% and the odds threshold $\theta$ to 2.
In preliminary experiments we found that with these settings an agent assigns more than 90\% of her own messages and about 5-20\% of the messages of any random other agent to be in her core interest.
However, as long as the decision on what is part of the core interest is consistent the overall model will behave very similar.

The actual act of rating a message as relevant  then depends on two probabilities $p_{\textit{core}}$ and $p_{\textit{peripheral}}$.
Not all messages of core interest will be considered relevant.
Neither will all messages of peripheral interest be considered irrelevant.
As we have no empirical data on which to base these probabilities will we keep them variable and explore the behaviour of the entire simulated community depending on these two probabilities.

\begin{table*}[tb]
\centering 
\caption{Parameters and their values in our simulation.}
\label{tab:parameters}
\begin{tabular}{llr}
\toprule
$|V|$		& Size of vocabulary, i.e. number of words & 10,000\\
$\alpha_{i}$	& Parameter vector entries in Dirichlet prior for the topic distributions of the agents & 0.01\\
$\beta_{i}$	& Parameter vector entries in Dirichlet prior for the term distributions of the topics & 0.001\\
$N$		& Network size, i.e. number of users & 10,000 \\
$m$		& Minimum number of friends per user & 5 \\
$k$		& Number of topics & 100\\
$l$		& Message length & 10\\
$e$		& Prior for the expected value of the number of messages a user creates	& 2.42\\
$p_{\textit{core}}$	& Probability to rate a message of core interest as relevant & variable\\
$p_{\textit{peripheral}}$	& Probability to rate a message of peripheral interest as relevant & variable\\
\emph{cov}	& Core topic coverage & 0.8 \\
$\theta$	& Threshold odds for core topic & 2.0 \\
$b$		& Cut off rank position in the filters & 20 \\
\bottomrule
\end{tabular}
\end{table*}

%%%%%%%%%%%%%%%%%%%%%%%%%%%%%%%%%%%%%%%%%%%%%%%%%%%%%%%%%%%%%%%%%%%%%%%%%%%%%%%%
\section{Personalisation Algorithms}
\label{sec:personalisation}
%%%%%%%%%%%%%%%%%%%%%%%%%%%%%%%%%%%%%%%%%%%%%%%%%%%%%%%%%%%%%%%%%%%%%%%%%%%%%%%%

The personalisation algorithms need to implement the two methods \textsc{filter} and \textsc{learnInterests} to work in the context of our simulation framework.
This essentially complies with the operation modes of these algorithm.
On the one hand, the algorithm needs to rank a set of messages in decreasing order of predicted interest for a user.
It operates as a filter if the resulting ranking is cut off at a certain position $b$ and all entries below that position are discarded and hidden from the user.
On the other hand it needs to interpret the interactions of the users with the messages which indicate relevance in order to derive the user's interests.

We implemented two prototypical, orthogonal personalisation algorithms. 
One algorithm is based on the content and one on the authors of a message. 
Both are based on established methods to predict if a given message with a specific content or from a specific author will be rated as relevant by a user.

\subsection{Content Based Personalisation}

A message is characterised by its individual words. 
This assumption underlies many information retrieval models.  
The terms are signals that either contribute or oppose  the probability of the message to be relevant. 
The information provided by the user via her interaction with the contents can help to understand which kind of signal is emitted by which term. 
This assumption underlies the concept of relevance feedback in search systems.

We implemented our content based personalisation algorithm  following the principles of relevance feedback in the binary independence model of probabilistic retrieval~\cite{J:JASIS:1976:RobertsonS}. 
This means, that for each document the user has seen, we keep track of whether or not she has rated it to be relevant. 
This relevance is directly mapped onto the single terms. 
And for each term $t_{i}$ we can estimate the probability of being relevant (using a random variable $R$) by:

\begin{equation}
p(R|t_{i}) = \frac{cr(t)+C_V}{cr(t)+cn(t)+2C_V}
\end{equation}

In this case $cr_{V}(t_i)$ is the number of times a user has rated a message containing term $t_{i}\in V$ as relevant.
Likewise, $cn_{V}(t_i)$ counts how often a message with this term has not been rated as relevant.
The constant $C_V$ is a smoothing parameter to overcome zero probabilities for unseen events.

Using these estimates allows to formulate a probability of relevance for a new incoming message. 
To this end the message $c$ is broken up into its individual terms. 
The probability of relevance is given by the product of the probabilities of relevance over the single terms:

\begin{equation}
p(R|c)=\prod_{t_i \in c} P(R|t_i)
\end{equation}

This probability can be computed for each incoming message. 
The messages are then sorted in decreasing order of probability following the probability ranking principle~\cite{J:Doc:1977:Robertson}.

\subsection{Author Based Personalisation}

Instead of focussing in the contents, an alternative paradigm is to consider the authors of a message as indicative signal for relevance. 
This is mainly motivated by the strong social component of online communities: it makes sense to consider that it is rather the news of specific people than specific contents the users want to see.
Again we can model this in a probabilistic way. 
Depending on how many times messages of one specific author $a$ have been seen and rated as relevant allows for estimating the probability of messages of this author to be relevant. 
Given that the author is the only signal considered we can then rank all incoming messages according to the probabilities:

\begin{equation}
p(R|a) = \frac{cr_A(a)+C_A}{cr_A(a)+cn_A(a) +2C_A}
\end{equation}

Here, $cr_{A}(a)$ counts the number of times a user has rated a message of author $a \in A$ as relevant, while $cn_{A}(a)$ counts how often the messages with this author were not relevant.
Again we use a smoothing constant $C_A$.

\section{Measuring Filter Bubble Effects}
\label{sec:metrics}

To measure the impact of personalisation we have to consider two kinds of effects.
On the one hand we have to measure the benefits of the personalisation on incoming content items, on the other hand we have to measure the filter bubble effect.
The improvement in the ranking of incoming items can be measured relatively easy by considering classical information retrieval metrics.
Precision measures the ratio of relevant items among the retrieved, i.e. here among the seen messages.
IR systems and also their personalisation components strive to improve precision.
Thus, an increase in precision at the individual level of single agents due to the use of a personalisation filter is sign for the benefits of the personalisation algorithms.
To observe this behaviour on the macro-level we use average precision over all agents.

To answer our questions of the strength of the filter bubble we need different metrics.
To the best of our knowledge, metrics for filter bubble effects have not been developed so far. 
However, there are clear signals for users being in a filter bubble: (1) a reduction of the active social context, (2) a reduction of the vocabulary a user perceives and (3) users perceives messages mainly or only from the fields of her core interests. 

To measure the active social context it is sufficient to count the number of distinct friends from which an agent perceives messages. 
If we denote the set of authors for a set of messages by $A(\textit{filtered}_{a}) = \{b \in \textit{Network} : \exists c \in \textit{filtered}_{a} : c.\textsc{author} = b\}$, then we can define the active social context of any given agent $a$ by:

\begin{equation}
ASC(a) = \frac{|A(\textit{filtered}_a)|}{|\textit{Network}.\textsc{friends}(a)|}
\end{equation}

This value measures the active social context for one agent only.
To observe macro effects, we aggregate and average the values over all agents in the simulation:

\begin{equation}
ASC = \frac{1}{|A|} \cdot \sum_{a \in A} ASC(a)
\end{equation}

However, even a low value of this metric does not necessarily indicate a poorer social context.
After all it is possible, that the context is small at each time step, but constantly changing with every iteration.
Figuratively speaking, a user might received messages always just from a small circle of her friends, but this circle is composed of different friends in each iteration.

Thus, we need to take the change rate of the social context  over a longer time into consideration.
This can be achieved by extending $ASC$ to operate on the authors seen over a longer time period (e.g. the last 10 iterations).
We will refer to this values as $ASC_{h}$.
Comparing $ASC$ to $ASC_{h}$ we can estimate the change rate.

The reduction of active vocabulary can be measured in an equivalent way.
We defined the according functions as $AV$ and $AV_{h}$.

The ratio $CR$ of core messages in the perceived messages finally can be computed similar to average precision.
If we define $\textit{core}_{a}$ as the set of all messages in $\textit{filtered}_{a}$ which are in the core interest of an agent, then we get:

\begin{equation}
CR = \frac{|\textit{core}_{a}|}{|\textit{filtered}_{a}|}
\end{equation}

As we interested in which parts of the online community are affected stronger by filter bubble effects we stratified the agents depending on their node degree.
We divided the agents in five strata.
The first stratum consisted of the agents with the highest node degree which covered 20\% of all network edges.
The next stratum contained all agents providing the next 20\% of the edges and so on.

%%%%%%%%%%%%%%%%%%%%%%%%%%%%%%%%%%%%%%%%%%%%%%%%%%%%%%%%%%%%%%%%%%%%%%%%%%%%%%%%
\section{Simulation Experiments}
\label{sec:experiments}
%%%%%%%%%%%%%%%%%%%%%%%%%%%%%%%%%%%%%%%%%%%%%%%%%%%%%%%%%%%%%%%%%%%%%%%%%%%%%%%%

As stated above we could estimate most parameters in our model from public statistics of online communities or by using values which have been reported in related work. 
One parameter we could not estimate are  the probabilities $p_{\textit{core}}$ and $p_{\textit{peripheral}}$ with which the users rate messages of their core or peripheral interest as relevant. 
Thus, we kept these parameters variable and observed how this user behaviour influences the filter bubble effect.

This lead to a setup where we ran for each personalisation algorithm independent simulations over 441 parameter settings to cover the parameter space of the probabilities $p_{\textit{core}}$ and $p_{\textit{peripheral}}$.
We iterated over these probabilities from 0 to 1 in steps of 0.05. 
Each simulation run was initiated with a new random network and random assignment of topic distributions to the users.
Then we ran the simulation for 100 iterations---a value which in preliminary experiments had shown that at this point the values become stable.

%%%%%%%%%%%%%%%%%%%%%%%%%%%%%%%%%%%%%%%%%%%%%%%%%%%%%%%%%%%%%%%%%%%%%%%%%%%%%%%%
\section{Results}
\label{sec:results}
%%%%%%%%%%%%%%%%%%%%%%%%%%%%%%%%%%%%%%%%%%%%%%%%%%%%%%%%%%%%%%%%%%%%%%%%%%%%%%%%

\begin{figure*}[tbh]
\centering
  \subfigure[Global Average]{
    \label{sfig:term:avgCR}    
    \includegraphics[width=\plotwidth\textwidth]{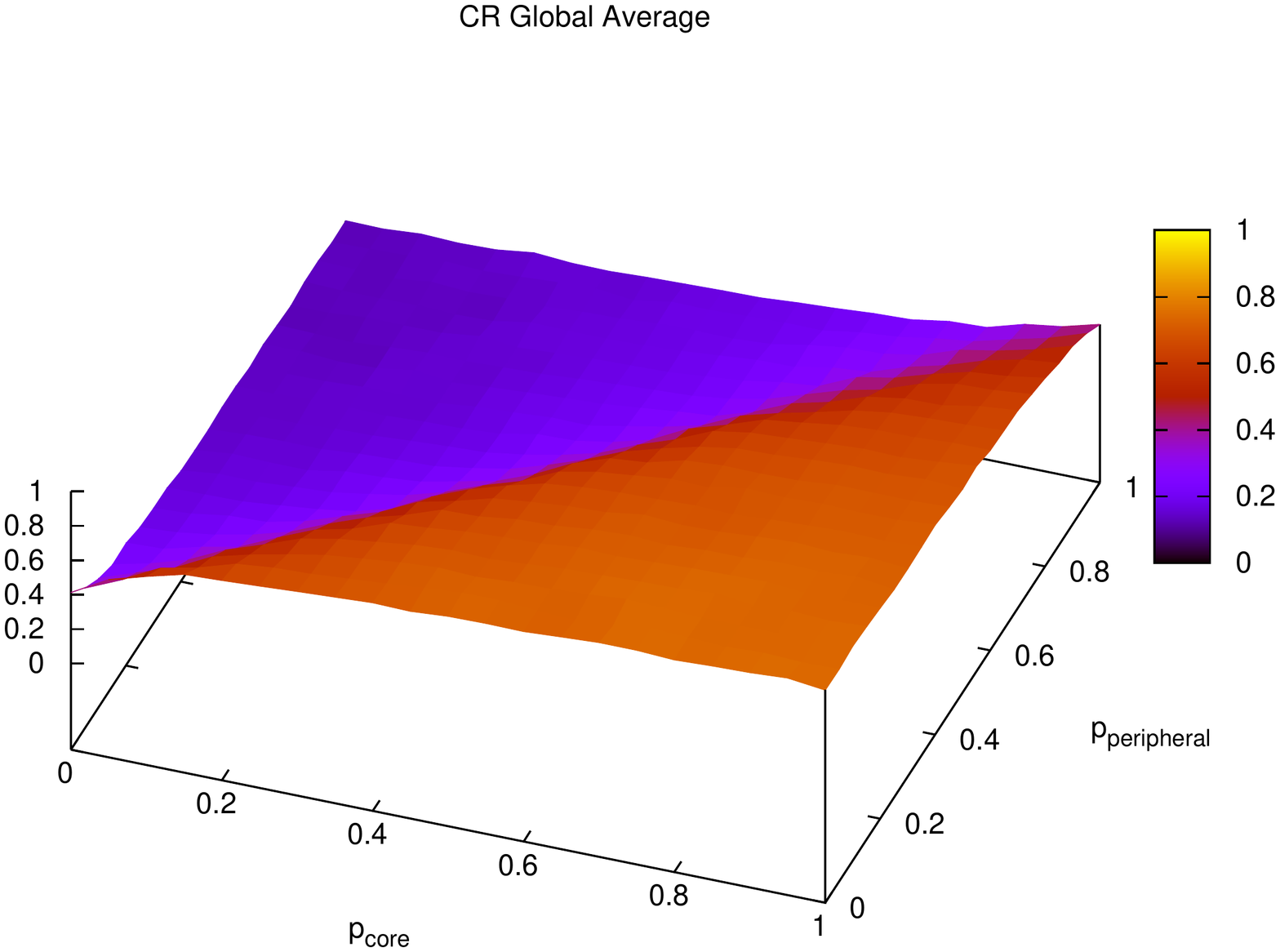} 
  }
  \subfigure[Top Stratum]{
    \label{sfig:term:t1CR}
    \includegraphics[width=\plotwidth\textwidth]{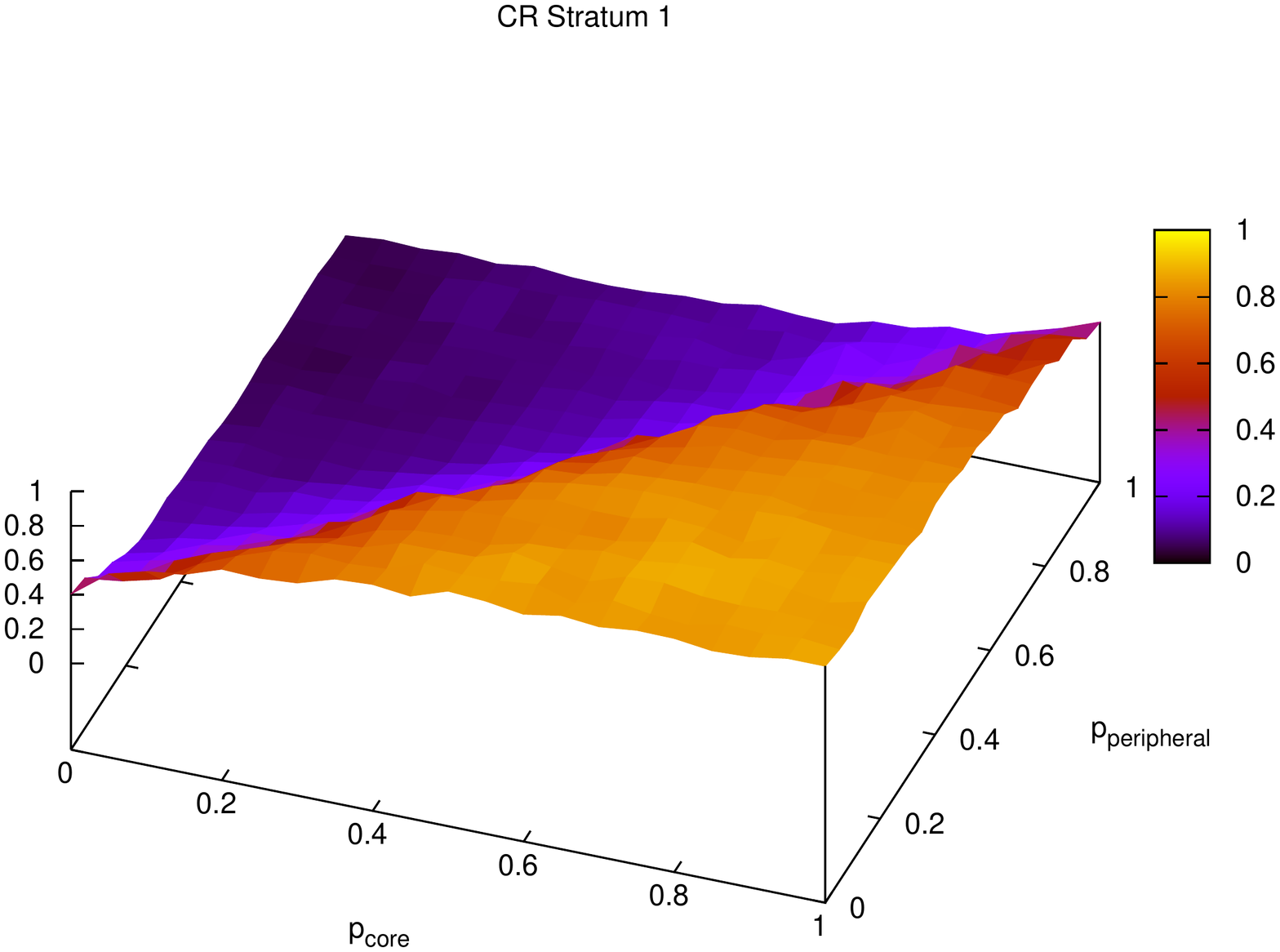} 
  }
\caption{$CR$ values for content based personalisation.}
\label{fig:termCR}
\end{figure*}

\begin{figure*}
\centering
  \subfigure[Global Average]{
    \label{sfig:author:avgCR}    
    \includegraphics[width=\plotwidth\textwidth]{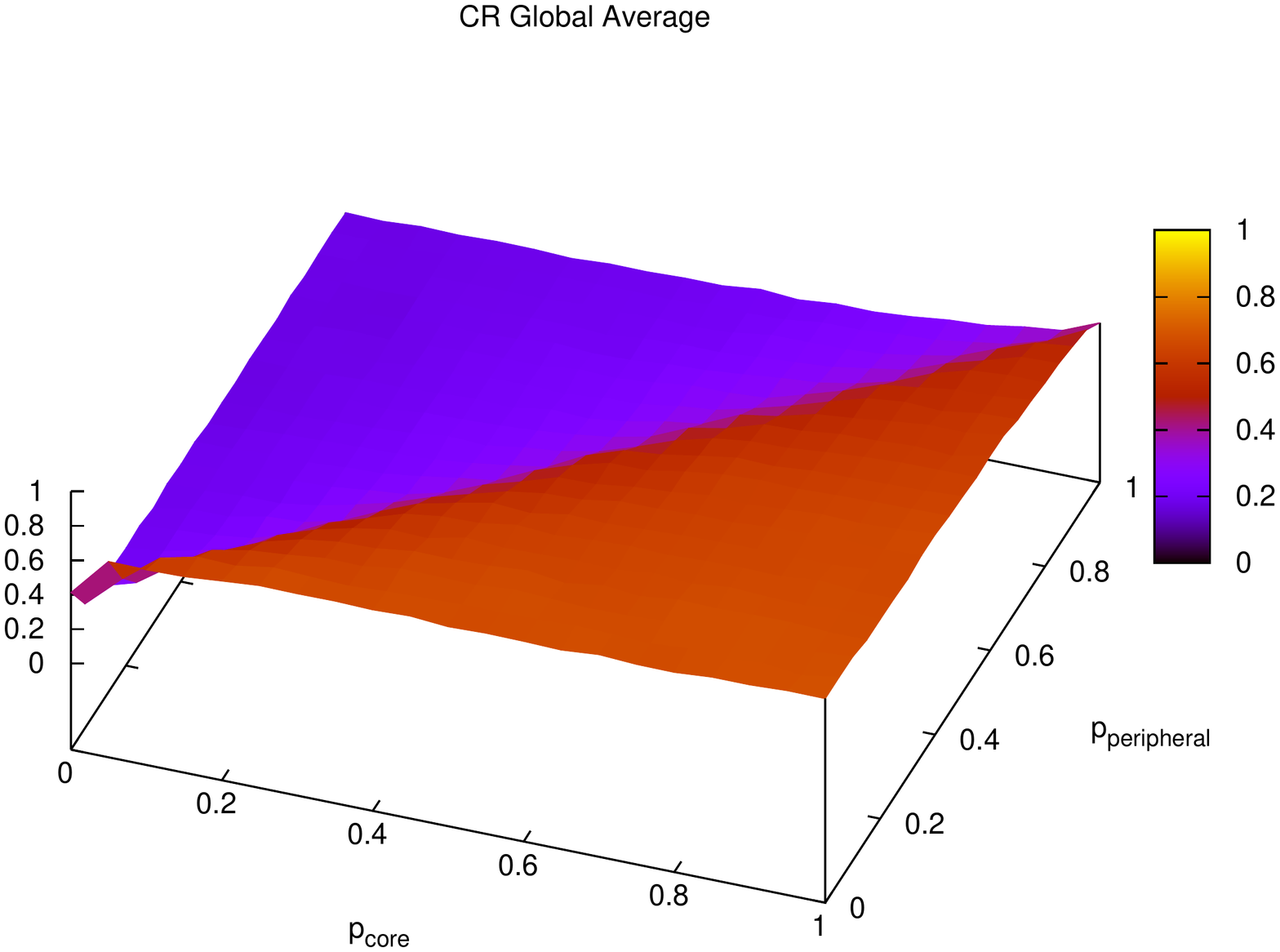} 
  }
  \subfigure[Top Stratum]{
    \label{sfig:author:t1CR}
    \includegraphics[width=\plotwidth\textwidth]{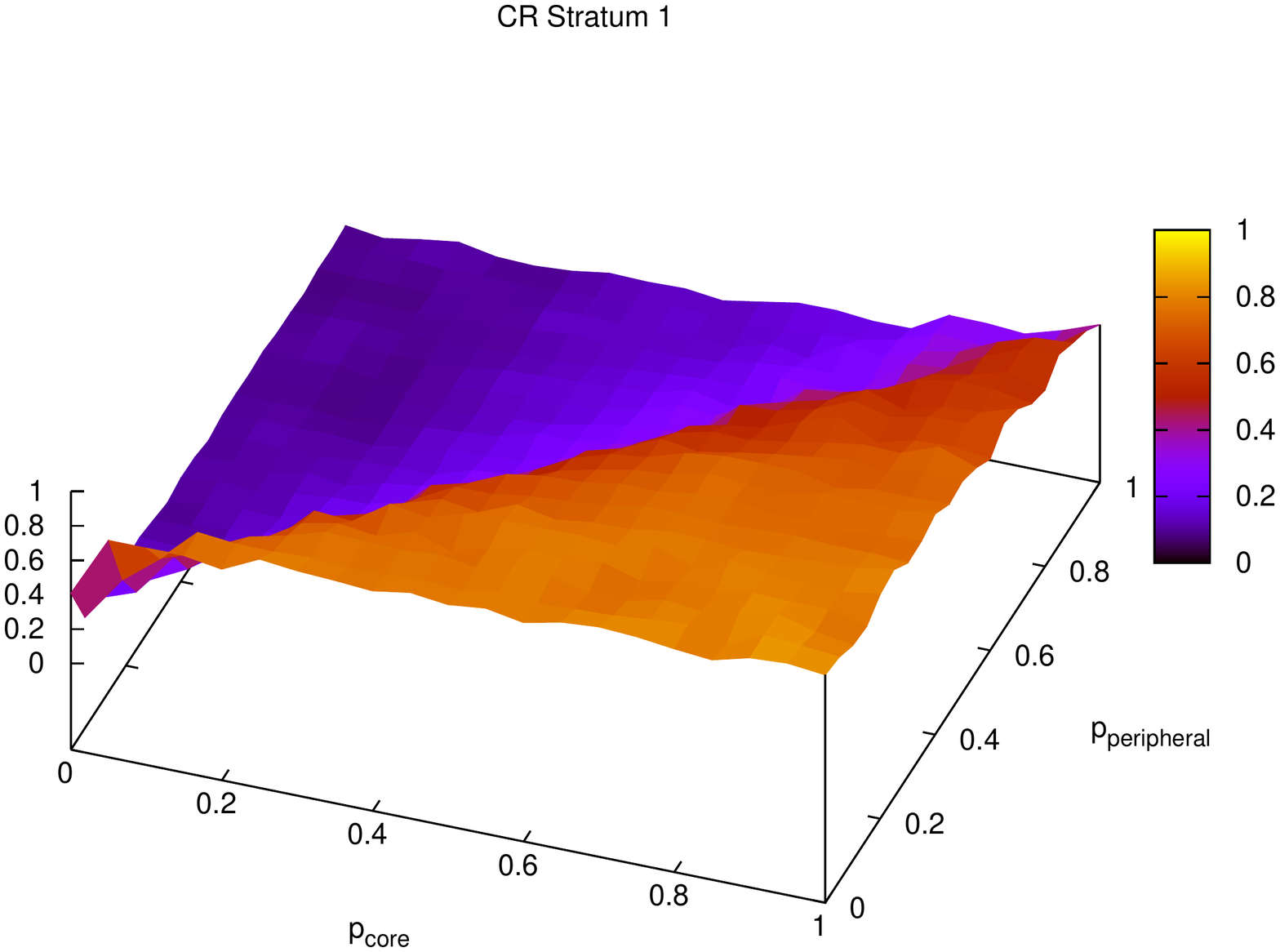} 
  }
\caption{$CR$ values for author based personalisation.}
\label{fig:authorCR}
\end{figure*}

Given that we iterate in small steps over a 2-dimensional parameter space we will present the result as three dimensional surface plots.
This allows to get a very good visual impression of the impact of the user behaviour (i.e. how likely they  rate messages of core and peripheral interest as relevant) on the filter bubble effect.

The first question we want to address is the impact on the ratio of messages of core interest the agents receive.
Figure~\ref{fig:termCR} shows the value of the $CR$ metric over the range of different settings for $p_{\textit{core}}$ and $p_{\textit{peripheral}}$ when operating with a content based personalisation.
The left Figure~\ref{sfig:term:avgCR} shows the average $CR$ for the entire online community.
We can see very clearly the diagonal ridge where the two probabilities are about the same value.
Above and below this ridge the $CR$ values form a plateau.
This means that the ratio of messages in the core interest field of an agent is higher if the probability for rating messages as relevant is higher in the class of core interest than in the category of peripheral messages.
This behaviour is somewhat expected as the personalisation filters learn what are the core interests of the users.
The plateau, however, indicates that the ratio of core interest messages does not grow arbitrarily high.
On the entire community, the values of $CR$ reach an average of at most 73\% if the users rate more core interest messages as relevant and at least 13\% if the users mainly rate peripheral interest messages as relevant.
The right plot in Figure~\ref{sfig:term:t1CR} shows the values for the top stratum of highly connected agents in the social network.
While the general behaviour is the same, the effects are much stronger.
The values in the lower plateau are set around 4\% and the high plateau values about 87\%.
This means that especially the well connected users are affected strongest by the filtering effects. 
They hardly get to see messages from outside their core field of interest.

Figure~\ref{fig:authorCR} shows the same plots for the personalisation algorithm based on message authors.
Again the general behaviour is very similar.
The effect, however, is slightly less strong.
The $CR$ values in the top stratum reaches only 80\%.

\begin{figure*}[tbh]
\centering
  \subfigure[Content base personalisation]{
    \label{sfig:term:t1ASC}    
    \includegraphics[width=\plotwidth\textwidth]{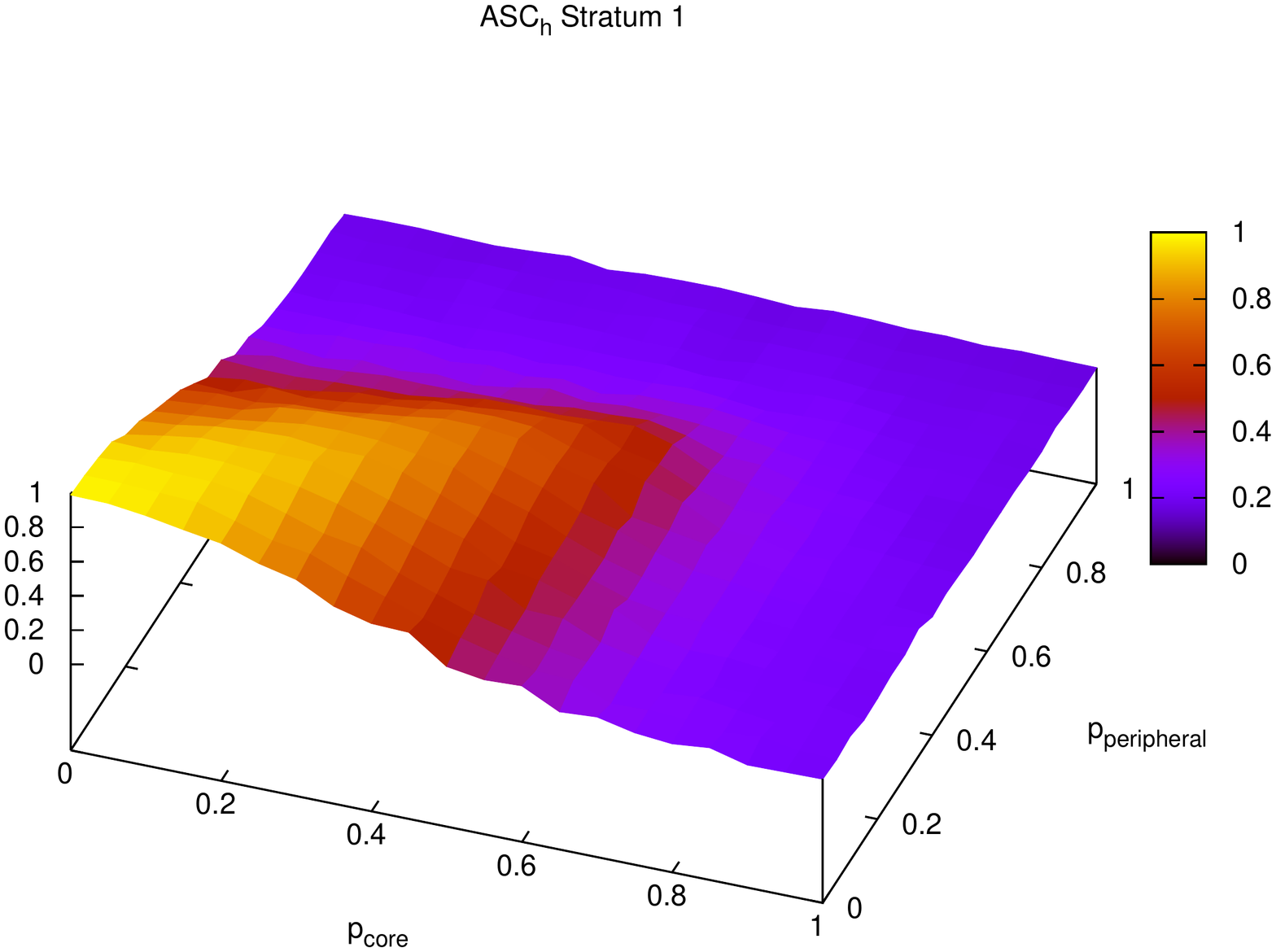} 
  }
  \subfigure[Author base personalisation]{
    \label{sfig:author:t1ASC}
    \includegraphics[width=\plotwidth\textwidth]{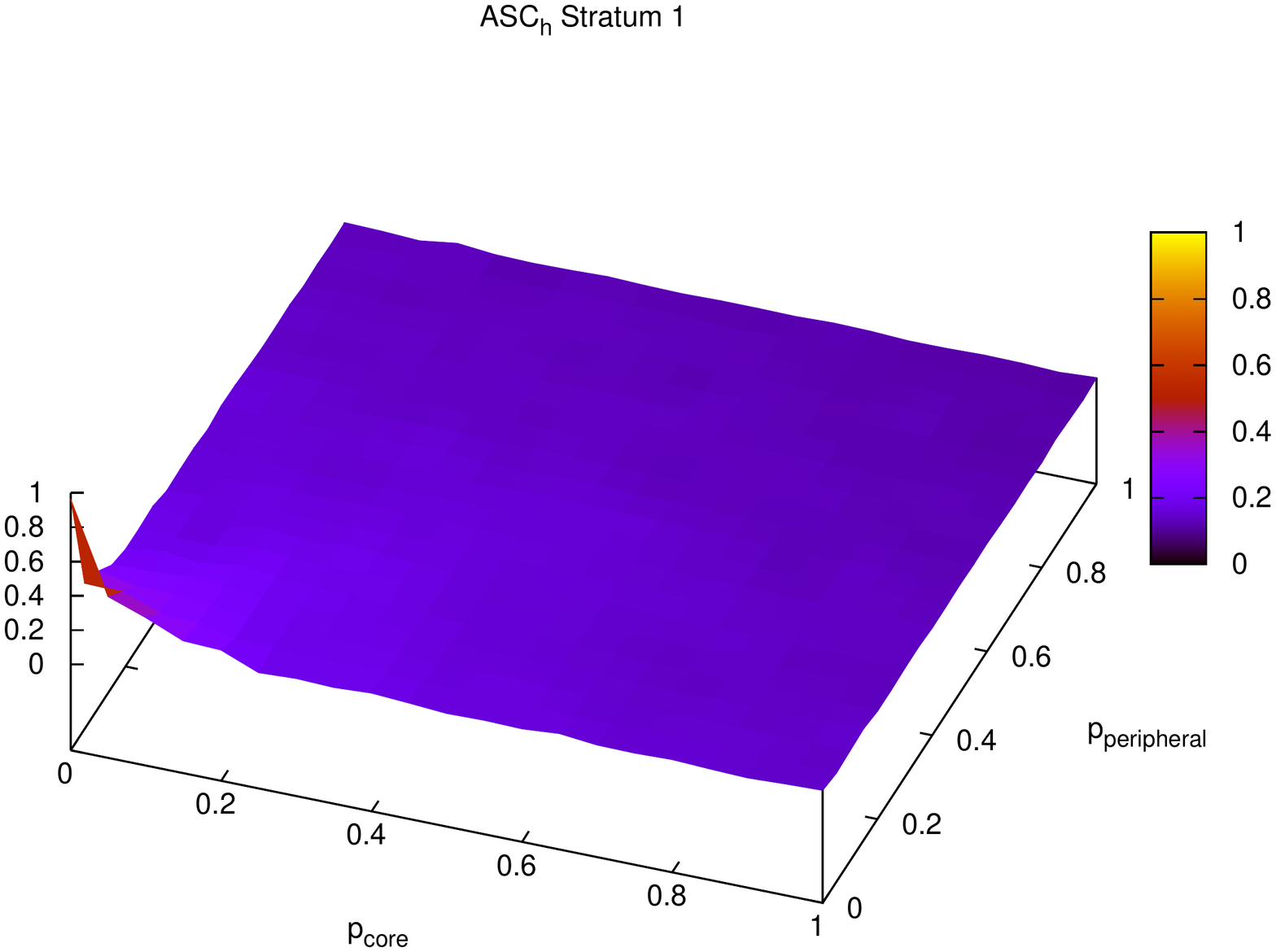} 
  }
\caption{Active social network $ASC_{h}$ for the top stratum.}
\label{fig:ascHist}
\end{figure*}

Let us look at the impact on the social network next.
In Figure~\ref{fig:ascHist} we compare the active social network ($ASC_{h}$) of the top stratum for content based (Figure~\ref{sfig:term:t1ASC}) and author based personalisation (Figure~\ref{sfig:author:t1ASC}).
Now we can see a clear difference between the two methods.
While the author based personalisation immediately restricts the number of people from which messages are received as soon as the agents give any feedback, the content based personalisation is not that extreme. 
The $ASC_{h}$ values are higher in general and the decline is much slower, when users give feedback.
Furthermore, as long as they do not excessively  rate messages as relevant (with a probability of more than 60\%) the active social context remains relatively broad.
This is also reflected in the change rate of the perceived authors.

\begin{figure*}
\centering
  \subfigure[Content base personalisation]{
    \label{sfig:term:t1AV}    
    \includegraphics[width=\plotwidth\textwidth]{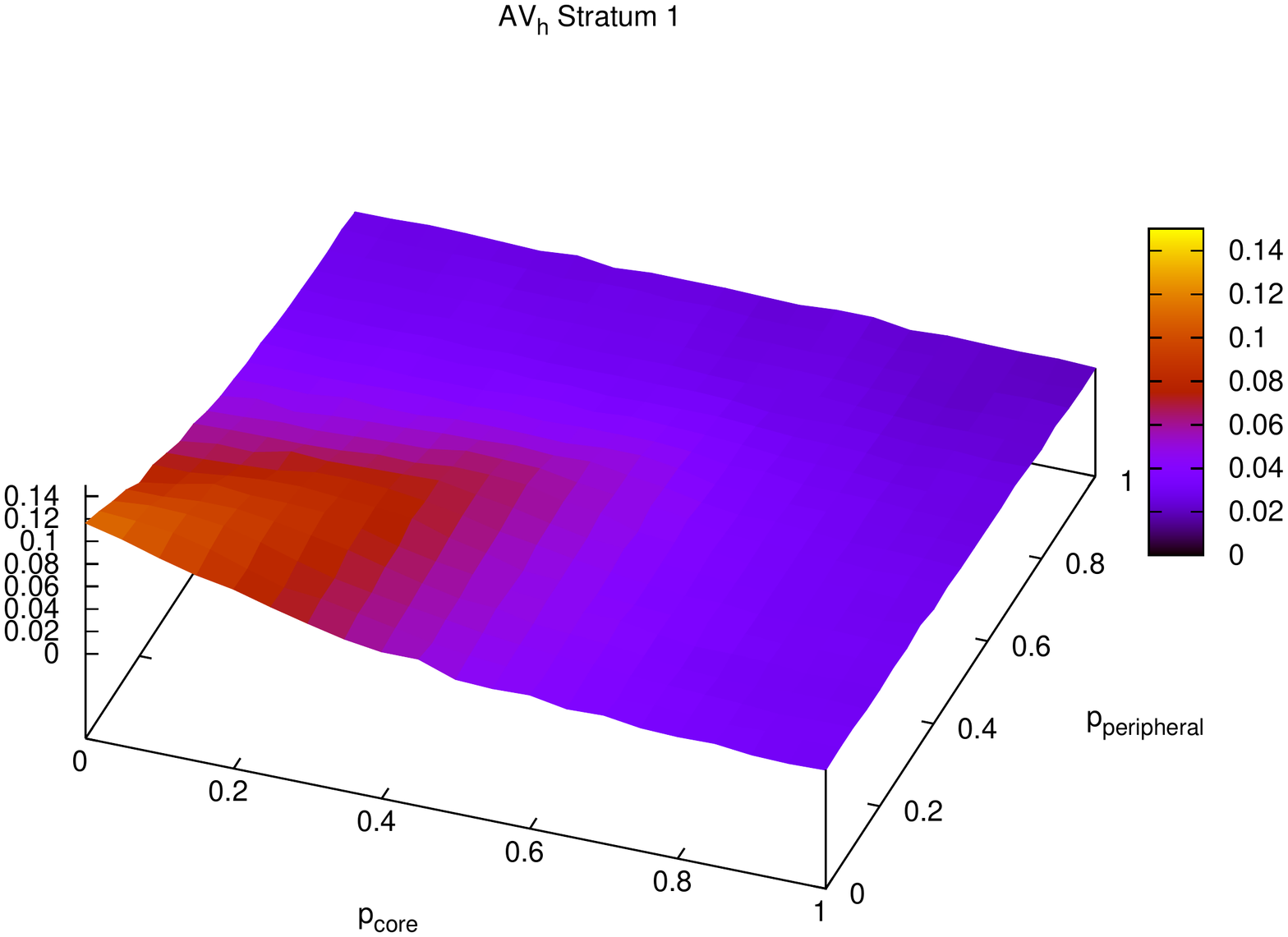} 
  }
  \subfigure[Author base personalisation]{
    \label{sfig:author:t1AV}
    \includegraphics[width=\plotwidth\textwidth]{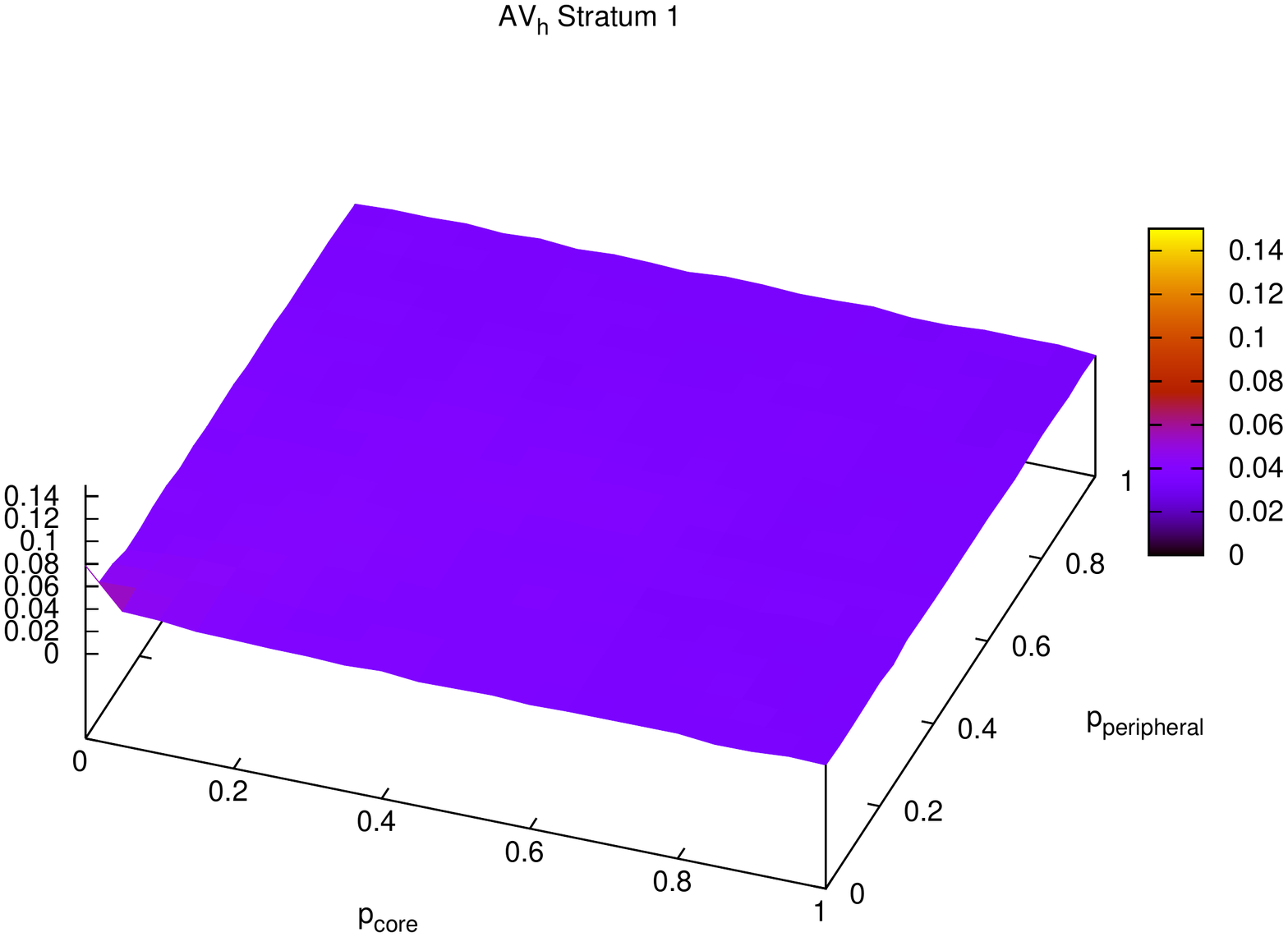} 
  }
\caption{Active vocabulary $AV_{h}$ for the top stratum.}
\label{fig:avHist}
\end{figure*}

A similar behaviour can be observed on the side of the vocabulary in Figure~\ref{fig:avHist}%
\footnote{Note the changed scale for the values. 
Given the size of the vocabulary, the few messages an agent receives after filtering can only cover a small fraction of the vocabulary anyway.}.
This is intriguing, as content based personalisation in the end relies on single terms. 
However, filtering messages based on the preferred terms restricts the overall perceived terms less than when filtering messages based on their author.

\begin{figure*}[tbh]
\centering
  \subfigure[Content base personalisation]{
    \label{sfig:term:t1p}    
    \includegraphics[width=\plotwidth\textwidth]{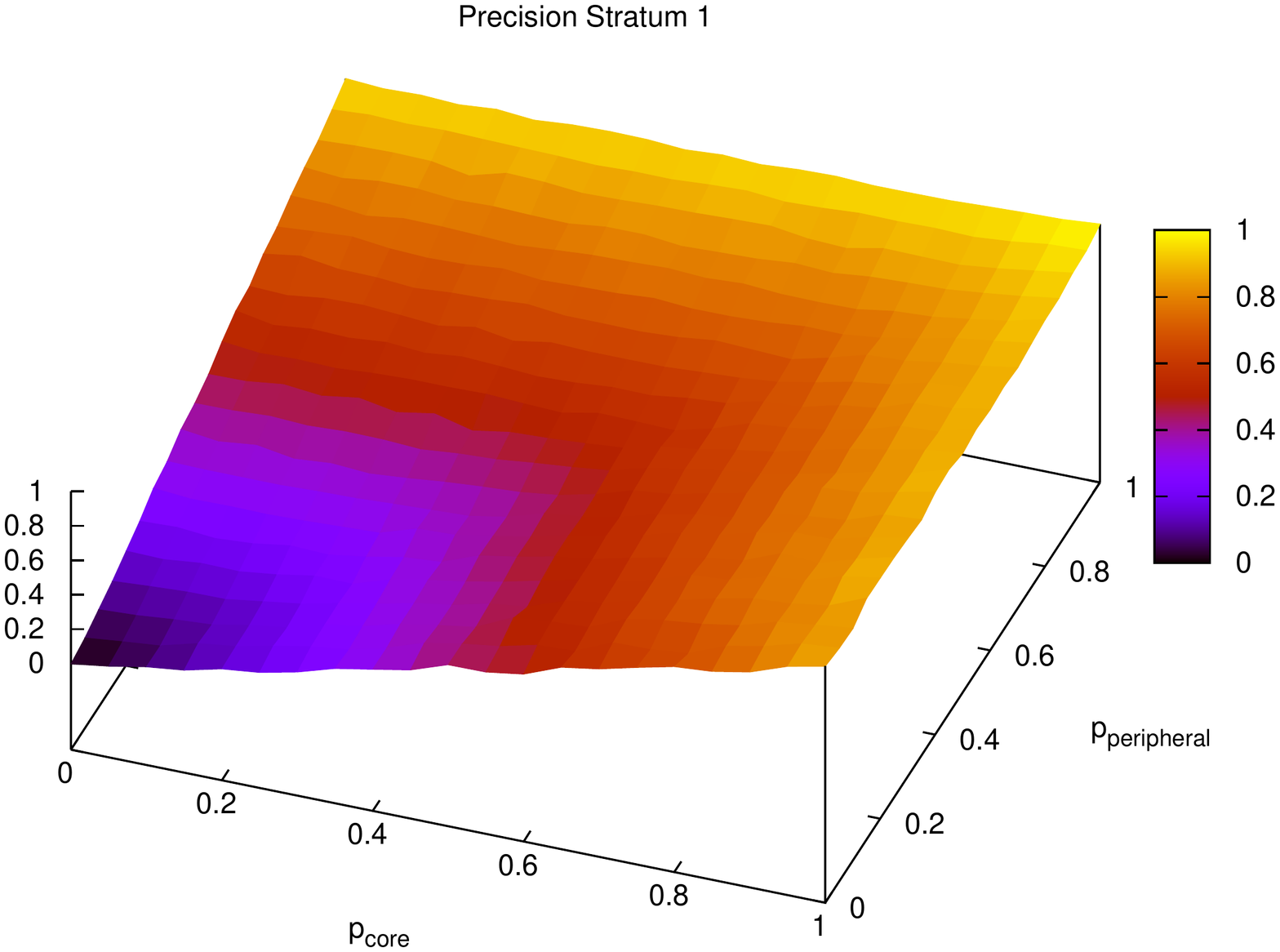} 
  }
  \subfigure[Author base personalisation]{
    \label{sfig:author:t1p}
    \includegraphics[width=\plotwidth\textwidth]{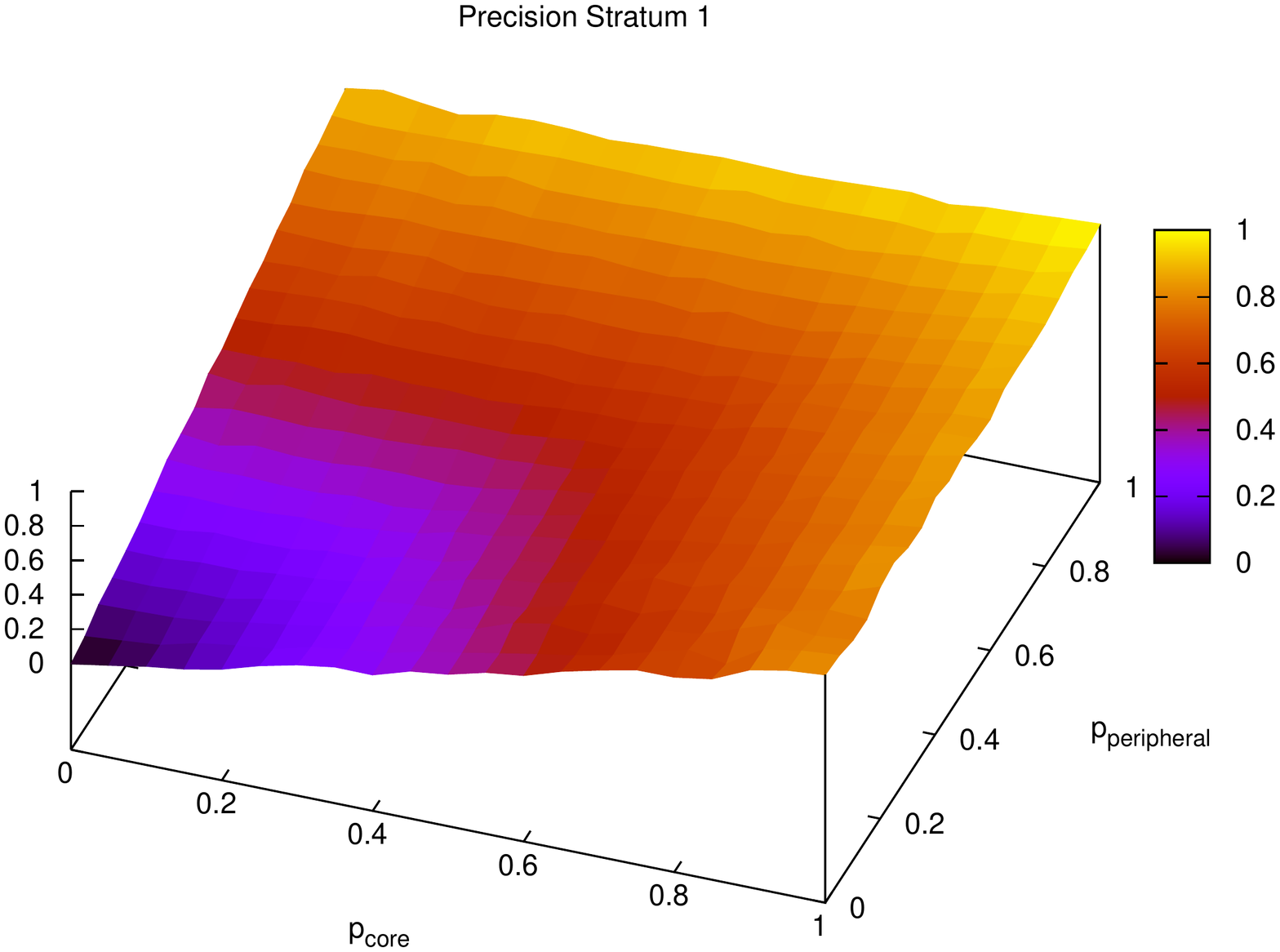} 
  }
\caption{Average precision for the top stratum.}
\label{fig:prec}
\end{figure*}

The positive aspects of the personalisation can be seen when looking at average precision in Figure~\ref{fig:prec}.
We can see that precision improves with the probability of the users giving feedback on what they consider relevant.
This is valid for both, the content and the author based personalisation.
However, the shape of these plots partially depends directly also on the probabilities $p_{\textit{core}}$ and $p_{\textit{peripheral}}$.
If those probabilities are close to zero, the agents rate no or hardly any message as relevant. 
Accordingly precision is low by definition.
However, if they tend to consider messages relevant at a high probability they will rate nearly everything as relevant and, thus, precision will automatically be very high.
The most interesting aspect of the plot are the regions where $p_{\textit{core}}$ and $p_{\textit{peripheral}}$ have very different values.
In these cases the algorithms really need to bring the right category of messages (core interest or peripheral interest) to the agent's attention such that they have the right basis of messages to rate with a higher probability as relevant.
Given the high values in these regions and the observations made for the $CR$ values, this is achieved by both approaches

%%%%%%%%%%%%%%%%%%%%%%%%%%%%%%%%%%%%%%%%%%%%%%%%%%%%%%%%%%%%%%%%%%%%%%%%%%%%%%%%
\section{Discussion}
\label{sec:discussion}
%%%%%%%%%%%%%%%%%%%%%%%%%%%%%%%%%%%%%%%%%%%%%%%%%%%%%%%%%%%%%%%%%%%%%%%%%%%%%%%%

Our simulation model allowed us to make some quite interesting observations on the macro level of online communities:

\begin{enumerate}

\item Filter bubble effects occur when using personalisation algorithms. 
We observed for both types of personalisation the effect, that users have a biased perception based on their own interests.
This happens as soon as they indicate a preference of their core interest by rating messages in this field as relevant with a different probability than all other messages. 
However, this after all is the aim of the personalisation algorithms. 
So this can be considered the central trade off between using personalisation for improving the user experience and not using personalisation for the sake of unbiased message perception.

Furthermore, on average the filter bubble effect is not that strong that a user cannot get out of the filter bubble any more. 
Users still receive messages from outside their core field of interest which technically allows them re-adjust the personalisation algorithms to potentially new interests.

\item Better connected users tend to be more prone to filter bubble effects than the average user. 
This can be easily explained with these users receiving far more messages.
Accordingly they are also likely to receive more relevant messages falling into field of their core interest and from which the personalisation algorithms can chose.  
The algorithms are simply less under pressure to also suggest less relevant messages.

However, this bears a certain risk as well: the well connected users represent hubs and are typically considered influential users. 
If their perception of the world is biased, this bias may affect also their social context and be propagated through the online community.

\item Author based personalisation causes a stronger filter bubble effect than content based personalisation under the aspects of active social context and active vocabulary. 
Our experiments showed a much stronger restriction of the active social network and perceived vocabulary when using the author based filtering algorithm. 
This is partially interesting as it might influence the potential to actually get out of the filter bubble.

\end{enumerate}

Given that both approaches behave similar under the aspect of focusing on the users interests and achieving comparable  precision values, the content based personalisation approach seems to have a lower impact on the filtering of the social network and vocabulary. 
Thus, one conclusion might be that content based personalisation seems the better option as it causes a less strong filter bubble effect.
However, this observation needs to be checked and confirmed with other personalisation algorithms.

%%%%%%%%%%%%%%%%%%%%%%%%%%%%%%%%%%%%%%%%%%%%%%%%%%%%%%%%%%%%%%%%%%%%%%%%%%%%%%%%
\section{Summary and Future Work}
\label{sec:summary}
%%%%%%%%%%%%%%%%%%%%%%%%%%%%%%%%%%%%%%%%%%%%%%%%%%%%%%%%%%%%%%%%%%%%%%%%%%%%%%%%

In this paper we proposed a model to measure the macro effects of personalisation in online communities with respect to filter bubble effects. 
We combined established topic and network models to build an agent-based simulation of an online community and investigated the impact of different personalisation algorithms on the diversity of topics, people and vocabulary the users perceive. 
We observed that filter bubble effects occur as soon as the users give some indications about the topics they are interested in. 
However beyond the filtering of topics we also observed that in our setting author based personalisation lead to stronger restrictions and thus stronger filtering effect on the social network and vocabulary.

In future work we will investigate other personalisation algorithms and consider flexible network structures.
Also networks formed on the basis of common interests of the agents will be considered.
One suitable candidate might be the social-circles model~\cite{J:PhysRev:2006:WhiteKTFW}.

\bibliographystyle{plain}
\bibliography{filterbubble}

\begin{thebibliography}{10}

\bibitem{J:Science:1999:BarabasiA}
Albert-L\'{a}szl\'{o} Barab\'{a}si and R\'{e}ka Albert.
\newblock Emergence of scaling in random networks.
\newblock {\em Science}, 286:509--512, 1999.

\bibitem{P:PLEAD:2012:Barocas}
Solon Barocas.
\newblock The price of precision: voter microtargeting and its potential harms
  to the democratic process.
\newblock In {\em Proceedings of the first edition workshop on Politics,
  elections and data}, PLEAD '12, pages 31--36, New York, NY, USA, 2012. ACM.

\bibitem{J:MLR:2003:BleiNJ}
David~M. Blei, Andrew~Y. Ng, and Michael~I. Jordan.
\newblock Latent dirichlet allocation.
\newblock {\em Journal of Machine Learning Research}, 3:993--1022, 2003.

\bibitem{J:JEP:2012:ComanH}
A.~Coman and W.~Hirst.
\newblock Cognition through a social network: The propagation of induced
  forgetting and practice effects.
\newblock {\em Journal of Experimental Psychology: General}, 141(2):321, 2012.

\bibitem{drosou2010search}
M.~Drosou and E.~Pitoura.
\newblock Search result diversification.
\newblock {\em ACM SIGMOD Record}, 39(1):41--47, 2010.

\bibitem{P:ICWSM:2010:HoggL}
Tad Hogg and Kristina Lerman.
\newblock Social dynamics of digg.
\newblock In {\em Proceedings of the Fourth International Conference on Weblogs
  and Social Media, ICWSM 2010, Washington, DC, USA, May 23-26, 2010}. The AAAI
  Press, 2010.

\bibitem{J:TIST:2012:LermanH}
Kristina Lerman and Tad Hogg.
\newblock Using stochastic models to describe and predict social dynamics of
  web users.
\newblock {\em ACM TIST}, 3(4):62, 2012.

\bibitem{Book:2011:Pariser}
E.~Pariser.
\newblock {\em The filter bubble: What the Internet is hiding from you}.
\newblock Penguin Press HC, 2011.

\bibitem{J:JASIS:1976:RobertsonS}
S.E. Robertson and K.~Sp{\"a}rck~Jones.
\newblock {Relevance weighting of search terms}.
\newblock {\em Journal of the American Society for information Science},
  27(3):129--146, 1976.

\bibitem{J:Doc:1977:Robertson}
Stephen Robertson.
\newblock {The Probability Ranking Principle in IR}.
\newblock {\em The Journal of documentation}, 33(4):294--304, 1977.

\bibitem{P:WebSci:2010:SchwagereitSS}
Felix Schwagereit, Sergej Sizov, and Steffen Staab.
\newblock Finding optimal policies for online communities with cosimo.
\newblock In {\em Proceedings of the WebSci10: Extending the Frontiers of
  Society On-Line, April 26-27th, 2010, Raleigh, NC: US}, 2010.

\bibitem{P:ECDL:2009:TomsM}
Elaine~G. Toms and Lori McCay-Peet.
\newblock Chance encounters in the digital library.
\newblock In {\em Proceedings of the 13th European conference on Research and
  advanced technology for digital libraries}, ECDL'09, pages 192--202, Berlin,
  Heidelberg, 2009. Springer-Verlag.

\bibitem{J:PhysRev:2006:WhiteKTFW}
Douglas~R. White, Nata{\v{s}}a Kej{\v{z}}ar, Constantino Tsallis, Doyne Farmer,
  and Scott White.
\newblock Generative model for feedback networks.
\newblock {\em Phys. Rev. E}, 73(1):016119, Jan 2006.

\bibitem{P:EC:2008:Wilkinson}
Dennis~M. Wilkinson.
\newblock Strong regularities in online peer production.
\newblock In {\em Proceedings of the 9th ACM conference on Electronic
  commerce}, EC '08, pages 302--309, New York, NY, USA, 2008. ACM.

\end{thebibliography}

\end{document}